\documentclass[aps,tightenlines,showpacs]{revtex4}%

\usepackage{amsfonts}
\usepackage{amsmath}
\usepackage{amssymb}
\usepackage{graphicx}

\begin{document}

\preprint{quant-ph/06}
\title[$K-$way negativities]{The $K-$way negativities as entanglement
measures }
\author{S. Shelly Sharma}
\email{shelly@uel.br}
\affiliation{Depto. de F\'{\i}sica, Universidade Estadual de Londrina, Londrina
86051-990, PR Brazil }
\author{N. K. Sharma}
\email{nsharma@uel.br}
\affiliation{Depto. de Matem\'{a}tica, Universidade Estadual de Londrina, Londrina
86051-990 PR, Brazil }
\thanks{}

\begin{abstract}
A classification of N-partite states, based on $K-$way $\left( 2\leq K\leq
N\right) $ negativities, is proposed. The $K-$way partial transpose with
respect to a subsystem is defined so as to shift the focus to $K-$way
coherences instead of $K$ subsystems of the composite system. For an
N-partite system the fraction of $K-$way negativity ($2\leq K\leq N$),
contributing to global negativity, is obtained. After minimizing $K-$way
negativities through local unitary qubit rotations, a combined analysis of $%
2-$way, $3-$way \ and global negativities is shown to provide distinct
measures of genuine tripartite, W-state like and bipartite entanglement, for
three qubit composite system. To illustrate the point, entanglement of three
qubit GHZ class states, W-class states, three boson state and noisy states
is analysed. In principle, a combined analysis of $K-$way negativities of
all the subsystems should lead to the sufficient condition for $K-$way
entanglement of the composite system. While pure N$-$partite entanglement of
a composite system is generated by N$-$way coherences, N$-$partite
entanglement in general can be present due to $\left( K<N\right) -$ way
coherences as well.
\end{abstract}

\maketitle


\section{Introduction}

Entanglement of quantum systems plays a central role in quantum information
processing. Quantum entanglement has made possible quantum state transport 
\cite{benn93,bouw97}, quantum communication over noisy channels \cite{brie98}%
, and quantum cryptography\cite{deut96}. As such characterization of quantum
entanglement is a fundamental issue. Several very interesting entanglement
measures have been devised \cite{benn96,pere96,horo96,woot98,wong01} for
analyzing pure and mixed state entanglement of multipartite systems.
Geometrical ideas have also been explored to understand entanglement in a
number of papers \cite{moss01,bern03,leva05}. Although bipartite
entanglement is well understood, many interesting aspects of multipartite
entanglement are still to be \ explored. Peres \cite{pere96} and the
Horedecki \cite{horo96,horo197,horo297} have shown a positive partial
transpose of a bipartite density operator to be a sufficient criterion for
classifying bipartite entanglement. Positive partial transpose has been
shown to be a necessary and sufficient condition \cite{horo96} for
separability of two-level bipartite systems. For higher dimensional systems
positive partial transpose is a necessary condition \cite{horo297}.
Negativity \cite{zycz98,vida00} based on Peres Horodecski PPT criterion has
been shown to be an entanglement monotone \cite{eise01,vida02,plen05}.
Negativity is an interesting concept being related to the eigenvalues of
partially transposed density matrix. For mixed states, use of negativity as
an entanglement measure is restricted to positive states. For an N-partite
composite system with three or more parties, it is found useful to split the
system in M parts with each party containing one or more sub-systems. A
hierarchic classification of arbitrary N-qubit mixed states, based on
separability and entanglement distillation properties of certain partitions,
has been given by D\"{u}r and Cirac \cite{dur00}. In this article, we
propose a classification of N-partite states based on $K-$way negativities,
where $2\leq K\leq N$. The principle underlying the definition of $K-$way
negativity for pure and mixed states of $N-$subsystems is NPT sufficient
condition. However, $K-$way partial transpose with respect to a subsystem is
defined so as to shift the focus to $K-$way coherences of the composite
system instead of $K$ subsystems of the composite system. In principle, a
combined analysis of $K-$way negativities for all the subsystems as well as
the negativities as defined in \cite{vida02}, should provide a measure of $%
K- $partite entanglement of the composite system. The fraction of $K-$way
negativity ($2\leq K\leq N$), contributing to negativity of partial
transpose of a given entangled state of an N-partite composite system, can
be calculated. For a three qubit system, $2-$way, $3-$way and global
negativities of all subsystems are shown to provide distinct measures of
genuine tripartite, W-state like and bipartite entanglement \ of the
composite system. Entanglement is invariant under local unitary rotations,
whereas, coherences are not so. A given state operator $\widehat{\rho }$ may
be mapped to another state operator $\widehat{\rho }_{\min }$, having
exactly the same entanglement as the state $\widehat{\rho }$ but having $K-$%
way negativities that are either invariant or increase under local unitary
rotations of qubits. Tripartite and bipartite entanglement measures for the
states $\widehat{\rho }$ and $\widehat{\rho }_{\min }$ may be written in
terms of fractional negativities of $3-$way and $2-$way partial transpose of
state $\widehat{\rho }_{\min }$. While pure N$-$ partite entanglement of a
composite system is generated by N$-$way coherences, N$-$partite
entanglement in general can be present due to $K-$way $\left( 2\leq
K<N\right) $ coherences as well.

We explain the notation used for computational basis states in section II.
The global negativity, which is twice the negativity as defined in ref. \cite%
{vida02} is discussed in section III. The $K-$way partial transpose of an $%
N- $partite state is defined in section IV, followed by the definition of $%
K- $way negativity with respect to a given subsystem in section V. To
clarify the definitions of section V, the construction of $2-$way and $3-$%
way partial transpose for three qubit states is given in section\ VI.
Section VII deals with the contribution of $K-$way negativities for a
particular subsystem to the global negativity of partially transposed state
operator. Minimization of $2-$way and $3-$way negativities for a set of one
parameter three qubit pure states is carried out in section VIII. Genuine
three qubit entanglement induced by $3-$way coherence and three qubit
entanglement induced by $2-$way coherence are also discussed, in the same
section. Sections IX and X deal with entanglement of three qubit two
parameter canonical states and three qubit noisy states, respectively.
Conclusions are given in section XI.

\section{The basis for N-partite system}

The Hilbert space, $C^{d}=C^{d_{1}}\otimes C^{d_{2}}\otimes ...\otimes
C^{d_{N}}$, associated with a quantum system composed of $N$ sub-systems, is
spanned by basis vectors of the form $\left\vert
i_{1}i_{2}...i_{N}\right\rangle ,$ where $i_{m}=0$ to $\left( d_{m}-1\right)
,$ and $m=1,...,N$. Here $d_{m}$ is the dimension of Hilbert space
associated with $m^{th}$ sub-system. We define a raising operator for the $%
m^{th}$ sub-system as $\sigma _{m}^{+}\left\vert i_{m}\right\rangle
=\left\vert i_{m}+1\right\rangle $ for $i_{m}<d_{m}-1$, and $\sigma
_{m}^{+}\left\vert d_{m}-1\right\rangle =0$. Using this definition, we may
rewrite%
\begin{align}
\left\vert i_{1}i_{2}...i_{N}\right\rangle & =(\sigma
_{1}^{+})^{i_{1}}(\sigma _{2}^{+})^{i_{2}}...(\sigma
_{N}^{+})^{i_{N}}\left\vert 0_{1}0_{2}...0_{N}\right\rangle  \notag \\
& =\left\vert j,\mu \right\rangle ,  \label{1}
\end{align}%
where $j=\sum\limits_{m=1}^{N}i_{m}\left(
\prod\limits_{n=1}^{m}d_{n-1}\right) $ with $d_{0}=1$, and $\mu
=\sum\limits_{m=1}^{N}i_{m}.$ The additional label $\mu $ counts the total
number of local raising operations needed to get the basis state $\left\vert
i_{1}i_{2}...i_{N}\right\rangle $ from the reference state $\left\vert
0_{1}0_{2}...0_{N}\right\rangle .$ For an $N$ qubit quantum system, $\sigma
_{m}^{+}$ is the usual spin raising operator for $m^{th}$ sub-system and the
label $\mu $ counts the total number of spins that are flipped to get the
basis vector $\left\vert i_{1}i_{2}...i_{N}\right\rangle $ from the
reference state $\left\vert 0_{1}0_{2}...0_{N}\right\rangle $.

Using this notation the state operator for N$-$partite composite system
operating on Hilbert space $C^{d}$ can be written as 
\begin{align}
\widehat{\rho} & =\sum_{(j,\mu)}\left\langle j,\mu\right\vert \widehat{\rho }%
\left\vert j,\mu\right\rangle \left\vert j,\mu\right\rangle \left\langle
j,\mu\right\vert  \notag \\
& +\sum_{\substack{ (j,\mu),(j\prime,\mu^{\prime})  \\ j\neq j\prime }}%
\left\langle j,\mu\right\vert \widehat{\rho}\left\vert j\prime,\mu^{\prime
}\right\rangle \left\vert j,\mu\right\rangle \left\langle
j^{\prime},\mu^{\prime}\right\vert .  \label{2}
\end{align}

\section{Global Negativity}

Consider a pure state or a mixed state with a density matrix which is known
to be positive. To measure overall entanglement of a subsystem $p$, we shall
use global negativity $N_{G}$, which is twice the negativity\ as
defined by Vidal and Werner \cite{vida02}. It is an entanglement measure for
classifying bipartite entanglement and is based on Peres-Hororedecki \cite%
{pere96,horo96} negative partial transpose (NPT) sufficient condition. The
global partial transpose of $\widehat{\rho }$ (Eq.(\ref{2})) with respect to
sub-system $p$ is defined as 
\begin{align}
\widehat{\rho }_{G}^{T_{p}}& =\sum_{(j,\mu )}\left\langle j,\mu \right\vert 
\widehat{\rho }\left\vert j,\mu \right\rangle \left\vert j,\mu \right\rangle
\left\langle j,\mu \right\vert  \notag \\
& +\sum_{\substack{ (j,\mu ),(j\prime ,\mu ^{\prime })  \\ j\neq ,j^{\prime
} }}\left\langle j,\mu \right\vert \widehat{\rho }\left\vert j\prime ,\mu
^{\prime }\right\rangle \left\vert j+\left( i_{p}^{\prime }-i_{p}\right)
\prod\limits_{n=1}^{p}d_{n-1},\mu +i_{p}^{\prime }-i_{p}\right\rangle  \notag
\\
& \left\langle j^{\prime }+\left( i_{p}-i_{p}^{\prime }\right)
\prod\limits_{n=1}^{p}d_{n-1},\mu ^{\prime }+i_{p}-i_{p}^{\prime }\right\vert
\label{5}
\end{align}%
The partially transposed matrix $\rho _{G}^{T_{p}}$ of an entangled state is
not positive definite. Global Negativity is defined as%
\begin{equation}
N_{G}^{p}=\left\Vert \widehat{\rho }_{G}^{T_{p}}\right\Vert
_{1}-1=2\sum\limits_{i}\left\vert \lambda _{i}^{-}\right\vert  \label{6}
\end{equation}%
where $\left\Vert \widehat{\rho }_{G}^{T_{p}}\right\Vert _{1}$ and $\lambda
_{i}^{-}$ are the trace norm and negative eigen values of $\rho _{G}^{T_{p}}$%
, respectively. For a state having $p^{th}$ subsystem separable, the value
of $N_{G}^{p}\ $is zero. Dimensionality of a composite entangled
system determines the maximum possible value of $N_{G}^{p}$. We
recall here that the partial transposition with respect to subsystem $p$
amounts to time reversal for the state of subsystem $p$. For an entangled
state this operation results in a state operator with negative eigenvalues.

In general, an $N-$partite system may not have pure $N-$partite entanglement
because $M$ ($1\leq M\leq N/2$ for $N$ even and $1\leq M\leq $ $(N-1)/2$ for 
$N$ odd) entangled subsystems are separable with respect to $N-M$
subsystems. In other words, the state operator can be written as%
\begin{equation}
\widehat{\rho }=\sum_{i}p_{i}\widehat{\rho }_{i}(N-M)\otimes \widehat{\rho }%
_{i}(M).  \label{7}
\end{equation}%
For a bipartite system $M=1$, as such, Peres Horodoski criteria separability
condition turns out to be a necessary and sufficient condition for
separability of the system. Since for a tripartite system $M=1$ as well, the
only condition for tripartite entanglement not to be present is that at
least one of the three subsystems is separable (that is one out of $N%
_{G}^{1}$, $N_{G}^{2}$, and $N_{G}^{3}$, is zero). On the
contrary, the necessary and sufficient condition for tripartite entanglement
to be present is that $\widehat{\rho }_{G}^{T}$ with respect to none of the
three subsystems is positive definite. That implies a non zero global
negativity ($N_{G}^{p}\neq 0$) for transposition with respect to $%
p^{th}$ sub-system, for all possible values of $p$. For a composite system
with four subsystems $M$ takes values $1$ and $2$. \ Hence there are two
types of conditions that forbid the existence of four party entanglement,
namely, the $4-$party entanglement is not present if $\rho _{G}^{T_{p}}$ is
positive for, at least, one of the four subsystems ($M=1$) or a pair of
subsystems ($M=2$). Similar considerations apply to systems with N greater
than four.

In the next two sections we fix the notation to write a\ $K-$way $\left(
2\leq K\leq N\right) $ partial transpose$\ $with respect to a single
subsystem and define $K-$way negativity. Once the presence of genuine $K-$%
way entanglement has been established, a measure of pure $K-$way
entanglement can be obtained from the $K-$way negativities of the composite
subsystem.

\section{$K-$way partial transpose}

A typical off diagonal matrix element of the state operator $\widehat{\rho}$
that involves a change of state of $K$ subsystems while leaving the state of 
$N-K$ sub-systems unaltered looks like $\left\langle
i_{1}i_{2}...i_{K,}i_{K+1,}...,i_{N}\right\vert \widehat{\rho}\left\vert
i_{1}^{\prime}i_{2}^{\prime}...i_{K,}^{\prime}i_{K+1,}...,i_{N}\right\rangle 
$. The set of $K$ distinguishable subsystems that change state while $N-K$
sub-systems do not, can be chosen in $D_{K}=\frac{N!}{(N-K)!K!}$ distinct
ways. If we represent the basis vectors obtained from $\left\vert
i_{1}i_{2}...i_{N}\right\rangle $ by changing the state of $\ K$ subsystems
by $\left\vert i_{1}^{\prime}i_{2}^{\prime}...i_{N}^{\prime},K\right\rangle
, $ where $K=0\ $to $N,$ the number of matrix elements of the type $%
\left\langle i_{1}i_{2}...i_{N}\right\vert \widehat{\rho}\left\vert
i_{1}^{\prime}i_{2}^{\prime}...i_{N,}^{\prime}K\right\rangle $ depends on $%
D_{K}$ and the dimensions $d_{1},d_{2}...d_{N}$ of the subsystems. For now
we stick to a simpler system of $N$ qubits, that is $%
d_{1}=d_{2}=...=d_{N}=2. $ Recalling that in the notation of section II the
matrix element $\left\langle i_{1}i_{2}...i_{N}\right\vert \widehat{\rho}%
\left\vert i_{1}^{\prime}i_{2}^{\prime }...i_{N,}^{\prime}K\right\rangle $
is written as $\left\langle j,\mu \right\vert \widehat{\rho}\left\vert
j\prime,\mu^{\prime}\right\rangle ,$ where $\mu=\sum\limits_{m=1}^{N}i_{m}$
and $\mu^{\prime}=\sum\limits_{m=1}^{N}i_{m}^{\prime}$, we have $%
K=\sum\limits_{m=1}^{N}\left\vert \left( i_{m}^{\prime}-i_{m}\right)
\right\vert .$ Alternatively, $K=\mu^{\prime}-\mu+2\sum\limits_{m=1,i>i^{%
\prime}}^{N}\left( i_{m}-i_{m}^{\prime}\right) $ and $\left\langle
j,\mu\right\vert \widehat{\rho}\left\vert j\prime ,\mu^{\prime}\right\rangle
=$ $\left\langle j,\mu\right\vert \widehat{\rho }\left\vert j\prime,\left(
\mu+K-2\sum\limits_{m=1,i>i^{\prime}}^{N}\left( i_{m}-i_{m}^{\prime}\right)
\right) \right\rangle $. Using the notation $\mu^{\prime}(K)=\mu+K-2\sum%
\limits_{m=1,i>i^{\prime}}^{N}\left( i_{m}-i_{m}^{\prime}\right) $, the
operator $\widehat{\rho}$ can be split up into parts labelled by $K$ with ($%
0\leq K\leq N).$ By rewriting the state operator as 
\begin{align}
\widehat{\rho} & =\sum_{_{(j,\mu),K=0}}\left\langle j,\mu\right\vert 
\widehat{\rho}\left\vert j,\mu\right\rangle \left\vert j,\mu\right\rangle
\left\langle j,\mu\right\vert
+\sum_{_{(j,\mu),(j\prime,\mu^{\prime}(1))}}\left\langle j,\mu\right\vert 
\widehat{\rho}\left\vert j\prime,\mu^{\prime }(1)\right\rangle \left\vert
j,\mu\right\rangle \left\langle j^{\prime},\mu^{\prime}(1)\right\vert  \notag
\\
& +\sum_{(j,\mu),(j\prime,\mu^{\prime}(2))}\left\langle j,\mu\right\vert 
\widehat{\rho}\left\vert j\prime,\mu^{\prime}(2)\right\rangle \left\vert
j,\mu\right\rangle \left\langle j^{\prime},\mu^{\prime}(2)\right\vert +... 
\notag \\
& +\sum_{_{(j,\mu),(j\prime,\mu^{\prime}(N))}}\left\langle j,\mu\right\vert 
\widehat{\rho}\left\vert j\prime,\mu^{\prime}(N)\right\rangle \left\vert
j,\mu\right\rangle \left\langle j^{\prime},\mu^{\prime}(N)\right\vert
=\sum_{K=0}^{N}\widehat{R}_{K},  \label{8}
\end{align}
where 
\begin{equation}
\widehat{R}_{K}=\sum_{(j,\mu),(j\prime,\mu^{\prime}(K))}\left\langle
j,\mu\right\vert \widehat{\rho}\left\vert
j\prime,\mu^{\prime}(K)\right\rangle \left\vert j,\mu\right\rangle
\left\langle j^{\prime},\mu^{\prime }(K)\right\vert,  \label{9}
\end{equation}
the terms containing matrix elements $\left\langle j,\mu\right\vert \widehat{%
\rho}\left\vert j\prime,\mu^{\prime}(K)\right\rangle $ that connect states
with a fixed value of $K$ are bunched together. We will refer to $\widehat{R}%
_{K}$ as the $K-$way coherence of $N-$qubit composite system. A three qubit
system, for example, may have $1-$way, $2-$way, and $3-$way coherence, $%
\widehat{R}_{0}$ being the diagonal part of $\ \widehat{\rho}$. Genuine $K-$%
partite entanglement cannot be generated if no $K-$way coherence is present.
But in general for $2\leq K<N$, the $K-$way coherence can result in
entanglement of more than $K$ parties.

Of the N sub-systems comprising the composite system, the $K-$way partial
transpose $\widehat{\rho }_{K}^{T_{p}}$ for the $p^{th}$ system is
constructed by partial transposition of $\ \widehat{R}_{K}$ with respect to
subsystem $p$ while leaving the rest of the state operator unchanged, that
is 
\begin{align}
\widehat{\rho }_{K}^{T_{p}}& =\sum_{K^{\prime }\neq K}^{N}\widehat{R}%
_{K^{\prime }}+\widehat{R}_{0}  \notag \\
& +\sum_{_{(j,\mu ),(j\prime ,\mu ^{\prime }(K))}}\left\langle j,\mu
\right\vert \widehat{\rho }\left\vert j\prime ,\mu ^{\prime
}(K)\right\rangle \left\vert j+\left( i_{p}^{\prime }-i_{p}\right) \left(
\prod\limits_{n=1}^{p}d_{n-1}\right) ,\mu +\left( i_{p}^{\prime
}-i_{p}\right) \right\rangle  \notag \\
& \left\langle j^{\prime }+\left( i_{p}-i_{p}^{\prime }\right) \left(
\prod\limits_{n=1}^{p}d_{n-1}\right) ,\mu ^{\prime }(K)+\left(
i_{p}-i_{p}^{\prime }\right) \right\vert .  \label{11}
\end{align}%
We recall here that the global partial transposition with respect to
subsystem $p$ amounts to time reversal for the state of subsystem $p$. For
an entangled state this operation results in a state operator with negative
eigenvalues. Taking $K-$way transpose with respect to $p^{th}$ subsystem
amounts to time reversal for the state of subsystem $p$ in $\widehat{R}_{K}$%
. In case $\sum_{K^{\prime }\neq K}^{N}\widehat{R}_{K^{\prime }}=0$, the
global and $K-$way partial transpose are the same. As such a $K-$way partial
transpose on $\widehat{\rho }$ may results in an operator with negative
eigenvalues iff $\widehat{R}_{K}\neq 0$ and $N_{G}^{p}\neq 0.$

\section{K-way Negativity}

The $K-$way negativity of subsystem $p$ to be calculated from $K-$way
partial transpose of matrix $\rho $ with respect to $p$, is defined as%
\begin{equation}
{N}_{K}^{p}=\left\Vert \widehat{\rho }_{K}^{T_{p}}\right\Vert _{1}-1,
\label{12}
\end{equation}%
where $\left\Vert \widehat{\rho }_{K}^{T_{p}}\right\Vert _{1}$ is the trace
norm of $\widehat{\rho }_{K}^{T_{p}}$ . The trace norm is calculated by
using the relation $\left\Vert \widehat{\rho }_{K}^{T_{p}}\right\Vert
_{1}=2\sum_{i}\left\vert \lambda _{i}^{-}\right\vert +1,$ $\lambda _{i}^{-}$
being the negative eigenvalues of matrix $\rho _{K}^{T_{p}}.$ The negativity 
$N_{K}^{p}$ depends on the $K-$ way coherence and is a measure of all
possible types of entanglement attributed to $K-$ way coherence.
Intuitively, for a system to have pure $N-$partite entanglement, it is
necessary that $N-$way coherences are non-zero for $p=1$ to $N$. On the
other hand, $N-$partite entanglement can also be generated by $(N-1)-$ way
coherence. For a three qubit system, maximally entangled tripartite GHZ
state is an example of pure tripartite entanglement involving $3-$way
coherence. On the other hand, maximally entangled W-state is a manifestation
of tripartite entanglement due to $2-$way coherences. Entanglement of a
subsystem is detected by global negativity $N_{G}^{p}$ ( for all possible
partitions of the system) and the hierarchy of negativities $N_{K}^{p}$ ($%
K=2,...N),$ calculated from $\widehat{\rho }_{K}^{T_{p}}$ associated with
sub-system $p=1$ to $N$. However, $N_{K}^{p}$ is not independent of the
global entanglement of the system. A necessary condition for pure $K-$%
partite entanglement to exist is that at least $K$ constituent systems have
non zero global as well as $K-$ way negativity.

\section{Constructing $\widehat{\protect\rho}_{G}^{T_{p}}$, $\widehat{%
\protect\rho}_{2}^{T_{p}}$and $\widehat{\protect\rho}_{3}^{T_{p}}$ for three
qubit states}

For a three qubit state, the global partial transpose with respect to first
qubit is constructed by transposing the state of qubit one while keeping the
composite state of qubits two and three unaltered. For the density operator 
\begin{align}
\widehat{\rho }& =\sum_{_{(j,\mu )}}\left\langle j,\mu \right\vert \widehat{%
\rho }\left\vert j,\mu \right\rangle \left\vert j,\mu \right\rangle
\left\langle j,\mu \right\vert  \notag \\
& +\sum_{_{\substack{ (j,\mu ),(j\prime ,\mu ^{\prime }(K))  \\ K=1,2,3}}%
}\left\langle j,\mu \right\vert \widehat{\rho }\left\vert j\prime ,\mu
^{\prime }(K)\right\rangle \left\vert j,\mu \right\rangle \left\langle
j^{\prime },\mu ^{\prime }(K)\right\vert  \notag \\
& =\sum_{K=0}^{3}\widehat{R}_{K}.  \label{20}
\end{align}%
where $j=\sum\limits_{m=1}^{3}i_{m}\left(
\prod\limits_{n=1}^{m}d_{n-1}\right) $ with $d_{0}=1$, and $\mu
=\sum\limits_{m=1}^{3}i_{m}$, the global partial transpose with respect to
qubit $p$ reads as%
\begin{align}
\widehat{\rho }_{G}^{T_{p}}=\sum_{_{(j,\mu )}}\left\langle j,\mu \right\vert 
\widehat{\rho }\left\vert j,\mu \right\rangle \left\vert j,\mu \right\rangle
\left\langle j,\mu \right\vert & +\sum_{_{\substack{ (j,\mu ),(j\prime ,\mu
^{\prime }(K))  \\ K=1,2,3}}}\left\langle j,\mu \right\vert \widehat{\rho }%
\left\vert j\prime ,\mu ^{\prime }(K)\right\rangle  \notag \\
\left\vert j+2^{p-1}\left( i_{p}^{\prime }-i_{p}\right) ,\mu +\left(
i_{p}^{\prime }-i_{p}\right) \right\rangle & \left\langle j^{\prime
}+2^{p-1}\left( i_{p}-i_{p}^{\prime }\right) ,\mu ^{\prime }(K)+\left(
i_{p}-i_{p}^{\prime }\right) \right\vert .  \label{20a}
\end{align}%
We write the $2-$way and $3-$way partial transpose with respect to qubit $p$
as%
\begin{align*}
\widehat{\rho }_{2}^{T_{p}}& =\sum_{_{(j,\mu )}}\left\langle j,\mu
\right\vert \widehat{\rho }\left\vert j,\mu \right\rangle \left\vert j,\mu
\right\rangle \left\langle j,\mu \right\vert \\
& +\sum_{_{\substack{ (j,\mu ),(j\prime ,\mu ^{\prime }(K))  \\ j\neq
,j^{\prime },K=1,3}}}\left\langle j,\mu \right\vert \widehat{\rho }%
\left\vert j\prime ,\mu ^{\prime }(K)\right\rangle \left\vert j,\mu
\right\rangle \left\langle j^{\prime },\mu ^{\prime }(K)\right\vert \\
& +\sum_{_{(j,\mu ),(j\prime ,\mu ^{\prime }(2))}}\left\langle j,\mu
\right\vert \widehat{\rho }\left\vert j\prime ,\mu ^{\prime }(2)\right\rangle
\\
& \left\vert j+2^{p-1}\left( i_{p}^{\prime }-i_{p}\right) ,\mu +\left(
i_{p}^{\prime }-i_{p}\right) \right\rangle \left\langle j^{\prime
}+2^{p-1}\left( i_{p}-i_{p}^{\prime }\right) ,\mu ^{\prime }(2)+\left(
i_{p}-i_{p}^{\prime }\right) \right\vert .
\end{align*}%
and%
\begin{align*}
\widehat{\rho }_{3}^{T_{p}}& =\sum_{_{(j,\mu )}}\left\langle j,\mu
\right\vert \widehat{\rho }\left\vert j,\mu \right\rangle \left\vert j,\mu
\right\rangle \left\langle j,\mu \right\vert \\
& +\sum_{_{\substack{ (j,\mu ),(j\prime ,\mu ^{\prime })  \\ K=1,2}}%
}\left\langle j,\mu \right\vert \widehat{\rho }\left\vert j\prime ,\mu
^{\prime }(K)\right\rangle \left\vert j,\mu \right\rangle \left\langle
j^{\prime },\mu ^{\prime }(K)\right\vert \\
& +\sum_{_{(j,\mu ),(j\prime ,\mu ^{\prime }(3))}}\left\langle j,\mu
\right\vert \widehat{\rho }\left\vert j\prime ,\mu ^{\prime }(3)\right\rangle
\\
& \left\vert j+2^{p-1}\left( i_{p}^{\prime }-i_{p}\right) ,\mu +\left(
i_{p}^{\prime }-i_{p}\right) \right\rangle \left\langle j^{\prime
}+2^{p-1}\left( i_{p}-i_{p}^{\prime }\right) ,\mu ^{\prime }(3)+\left(
i_{p}-i_{p}^{\prime }\right) \right\vert .
\end{align*}

Alternatively, the matrix representation may be used to show the action of $%
3-$way map $T_{3}^{1}$ defined as $T_{3}^{1}:\widehat{\rho }\rightarrow 
\widehat{\rho }_{3}^{T_{1}}$. The matrix $T_{3}^{1}$ acts only on those
matrix elements for which $K=\mu ^{\prime }-\mu
+2\sum\limits_{m=1,i>i^{\prime }}^{3}\left( i_{m}-i_{m}^{\prime }\right) =3$%
. We can arrange these matrix elements in a single column with eight rows.
Operation of the matrix $T_{3}^{1}$ on the column looks like%
\begin{equation}
\left[ 
\begin{array}{cccccccc}
0 & 1 & 0 & 0 & 0 & 0 & 0 & 0 \\ 
1 & 0 & 0 & 0 & 0 & 0 & 0 & 0 \\ 
0 & 0 & 0 & 1 & 0 & 0 & 0 & 0 \\ 
0 & 0 & 1 & 0 & 0 & 0 & 0 & 0 \\ 
0 & 0 & 0 & 0 & 0 & 1 & 0 & 0 \\ 
0 & 0 & 0 & 0 & 1 & 0 & 0 & 0 \\ 
0 & 0 & 0 & 0 & 0 & 0 & 0 & 1 \\ 
0 & 0 & 0 & 0 & 0 & 0 & 1 & 0%
\end{array}%
\right] \left[ 
\begin{array}{c}
\rho _{000,111} \\ 
\rho _{100,011} \\ 
\rho _{010,101} \\ 
\rho _{110,001} \\ 
\rho _{001,110} \\ 
\rho _{101,010} \\ 
\rho _{011,100} \\ 
\rho _{111,000}%
\end{array}%
\right] =\allowbreak \left[ 
\begin{array}{c}
\rho _{100,011} \\ 
\rho _{000,111} \\ 
\rho _{110,001} \\ 
\rho _{010,101} \\ 
\rho _{101,010} \\ 
\rho _{001,110} \\ 
\rho _{111,000} \\ 
\rho _{011,100}%
\end{array}%
\right] .  \label{22}
\end{equation}%
The full $\rho $ matrix, written as a column, has $64$ rows. The full map $%
T_{3}^{1}$ leaves the remaining $56$ matrix elements without any change.
Similarly, one can construct the maps $T_{3}^{2,3}$ and $T_{2}^{1,2,3}$. The
eigenvalues of partially transposed matrices $\rho _{G}^{T_{p}}$, $\rho
_{2}^{T_{p}}$, and $\rho _{3}^{T_{p}}$, where $p=1,2,$ and $3$, are used to
calculate the corresponding negativities.

\section{Contribution of K-way negativity to Global negativity}

Global negativity with respect to a subsystem can be written as a sum of
partial $K-$way negativities. The global transpose with respect to subsystem 
$p$, written in its eigen basis is given by 
\begin{equation}
\widehat{\rho }_{G}^{T_{p}}=\sum\limits_{i}\lambda _{i}^{G+}\left\vert \Psi
^{G+}\right\rangle \left\langle \Psi _{i}^{G+}\right\vert
+\sum\limits_{i}\lambda _{i}^{G-}\left\vert \Psi _{i}^{G-}\right\rangle
\left\langle \Psi _{i}^{G-}\right\vert ,  \label{1n}
\end{equation}%
where $\lambda _{i}^{G+}$and $\left\vert \Psi ^{G+}\right\rangle $ ($\lambda
_{i}^{G-}$and $\left\vert \Psi ^{G-}\right\rangle $) are the positive
(negative) eigenvalues and eigenvectors, respectively. The trace norm of $%
\widehat{\rho }_{G}^{T_{p}}$ is 
\begin{eqnarray}
\left\Vert \widehat{\rho }_{G}^{T_{p}}\right\Vert _{1}
&=&\sum\limits_{i}\left\langle \Psi _{i}^{G+}\right\vert \widehat{\rho }%
_{G}^{T_{p}}\left\vert \Psi _{i}^{G+}\right\rangle
-\sum\limits_{i}\left\langle \Psi _{i}^{G-}\right\vert \widehat{\rho }%
_{G}^{T_{p}}\left\vert \Psi _{i}^{G-}\right\rangle  \notag \\
&=&Tr\left( \widehat{\rho }_{G}^{T_{p}}\right) -2\sum\limits_{i}\left\langle
\Psi _{i}^{G-}\right\vert \widehat{\rho }_{G}^{T_{p}}\left\vert \Psi
_{i}^{G-}\right\rangle \text{.}
\end{eqnarray}%
Using $Tr\left( \widehat{\rho }_{G}^{T_{p}}\right) =1,$ the negativity of $%
\widehat{\rho }_{G}^{T_{p}}$ is given by 
\begin{equation}
{N}_{G}^{p}=-2\sum\limits_{i}\left\langle \Psi _{i}^{G-}\right\vert 
\widehat{\rho }_{G}^{T_{p}}\left\vert \Psi _{i}^{G-}\right\rangle
=-2\sum\limits_{i}\lambda _{i}^{G-}\text{.}  \label{2n}
\end{equation}%
The global transpose with respect to subsystem $p$, may also be rewritten as 
\begin{equation}
\widehat{\rho }_{G}^{T_{p}}=\sum\limits_{K=2}^{N}\widehat{\rho }%
_{K}^{T_{p}}-(N-2)\widehat{\rho }.  \label{3n}
\end{equation}

Substituting Eq. (\ref{3n}) in Eq. (\ref{2n}), and recalling that $\widehat{%
\rho }$ is a positive operator with trace one, we get%
\begin{equation}
-2\sum\limits_{i}\lambda
_{i}^{G-}=-2\sum\limits_{K=2}^{N}\sum\limits_{i}\left\langle \Psi
_{i}^{G-}\right\vert \widehat{\rho }_{K}^{T_{p}}\left\vert \Psi
_{i}^{G-}\right\rangle .
\end{equation}%
Defining partial $K-$way negativity $E_{K}^{p}$ ($K=2$ to $N$) as%
\begin{equation}
E_{K}^{p}=-2\sum\limits_{i}\left\langle \Psi _{i}^{G-}\right\vert \widehat{%
\rho }_{K}^{T_{p}}\left\vert \Psi _{i}^{G-}\right\rangle \text{,}
\end{equation}%
we may split the global negativity for qubit $p$ as%
\begin{equation}
N_{G}^{p}=\sum\limits_{K=2}^{N}E_{K}^{p}\text{.}
\end{equation}%
The contribution $E_{K}^{p}$ \ depends on $\left\Vert \rho
_{K}^{T_{p}}\right\Vert _{1}$, the ratio $E_{K}^{p}/N_{K}^{p}$ being the
fraction of $K-$way negativity contributing to $N_{G}^{p}$.

\begin{figure}[t]
\centering \includegraphics[width=3.75in,height=5.0in,angle=-90]{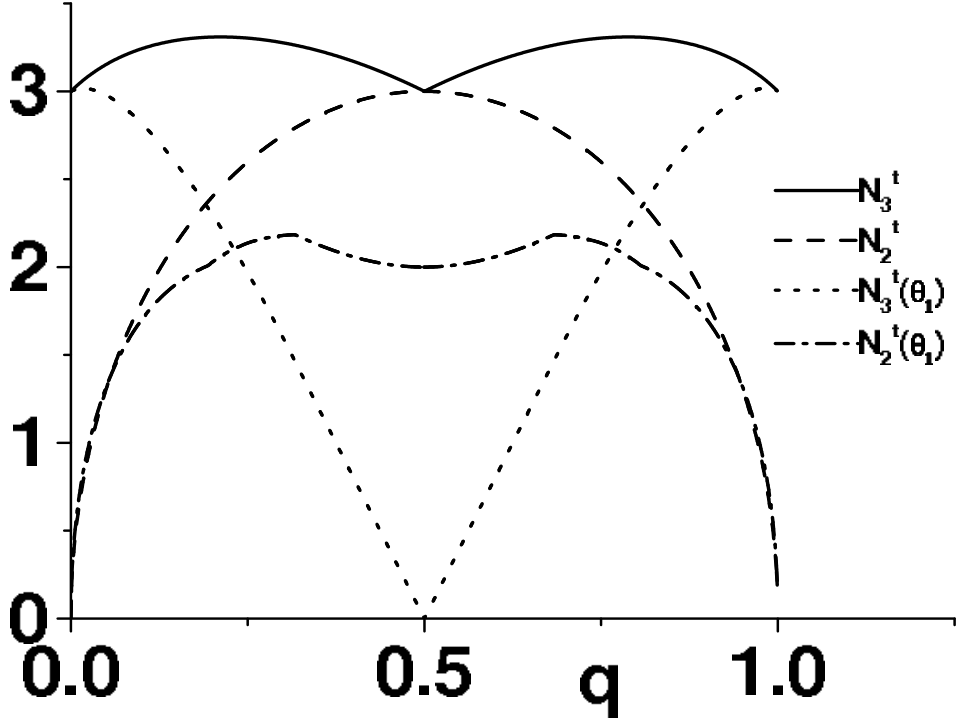}
\caption{Total negativities $N_{3}^{t}$, $N_{2}^{t}$, $N_{3}^{t}(\protect%
\theta _{1})$ and $N_{2}^{t}(\protect\theta _{1})$ versus parameter $q$ for
the pure state $\hat{\protect\rho}_{1}$.}
\label{fig1}
\end{figure}

The Peres-Horodecki criterion is the most useful sufficient condition for
checking the separability of a subsystem in a bipartite split of composite
quantum system. The sufficient condition for an $N-$partite pure state not
to have genuine $N-$partite entanglement is that at least one of the global
negativities is zero that is $N_{G}^{p}=0,$ where $p$ is one of the
subsystems or one part of a bipartite split of the composite system.
Consider an $N-$partite system in a pure state, $\widehat{\rho }=\widehat{%
\rho }_{0}+\widehat{\rho }_{1}+...+\widehat{\rho }_{N}$. Recalling that $%
N_{G}^{p}=\sum\limits_{K=2}^{N}E_{K}^{p}$ , the separability of subsystem $p$
implies that $E_{K}^{p}=0,$ or $\widehat{\rho }_{K}^{T_{p}}\geqslant 0$, for 
$K=2$ to $N$. In general, for a system having only genuine $K-$ partite
entanglement, $N_{G}^{p}=0$ for $N-K$ of the subsystems. In addition, the
lowest of the partial non zero $K-$way negativities measures $K-$partite
entanglement, the same being a collective property of $K-$subsystems.

\section{Local unitary operations and K-way coherences}

An important point to note is that the trace norm satisfies%
\begin{equation*}
\left\Vert \widehat{\rho }_{G}^{T_{p}}\right\Vert _{1}\leq
\sum\limits_{K=2}^{N}\left\Vert \widehat{\rho }_{K}^{T_{p}}\right\Vert
_{1}-(N-2)\left\Vert \widehat{\rho }\right\Vert _{1},
\end{equation*}%
or in terms of negativities%
\begin{equation*}
{N}_{G}^{p}\leq \sum\limits_{K=2}^{N}{N}_{K}^{p}.
\end{equation*}%
The trace norm $\left\Vert \widehat{\rho }_{K}^{T_{p}}\right\Vert _{1}$ is
not invariant under local unitary rotations, where as $\left\Vert \widehat{%
\rho }_{G}^{T_{p}}\right\Vert _{1}$ is invariant. For a given state $%
\widehat{\rho }$ , there exists a set of operators $\widehat{\rho }%
_{K}^{T_{p}}$ for which $\sum\limits_{p=1}^{N}{N}_{K}^{p}$ \ has the
lowest value, for all $K$. The state $\widehat{\rho }_{\min }$ that yields
transposed $K-$way operators with the lowest valued norm can be obtained
from $\widehat{\rho }$ by applying local unitary operations on constituent
sub-systems. We conjecture that for the state having $\sum\limits_{p=1}^{N}%
{N}_{K}^{p}$ \ at it's minimum,\ $E_{K}^{p}$ measures the $K-$way
entanglement of qubit $p$. Recalling that N-way entanglement is a collective
property of $N-$qubits,\ $\min (E_{N}^{1},E_{N}^{2},...,E_{N}^{N})$ is a
measure of genuine $N-$way entanglement of qubit $p$.

\begin{figure}[t]
\centering \includegraphics[width=3.75in,height=5.0in,angle=-90]{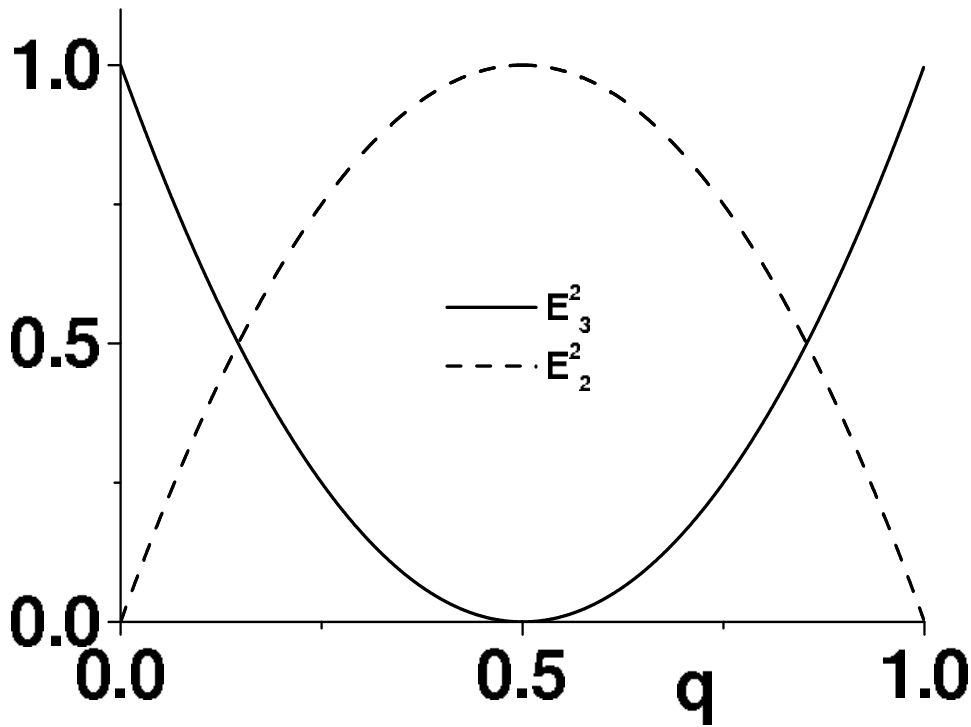}
\caption{Genuine tripartite entanglement, $E_{3}$ and bipartite entanglement 
$E_{2}^{2}(\protect\theta _{1})$ as function of parameter $q$ for the pure
state $\hat{\protect\rho}_{1}$.}
\label{fig2}
\end{figure}

It is well known that the set of states that can be transformed into each
other by local unitary operations lie on the same orbit and have the same
entanglement as the canonical state expressed in terms of the minimum number
of independent vectors \cite{cart99}. Construction of pure three qubit
canonical state has been given by Acin et al \cite{acin01}. It is easily
verified that for the states reducible to the canonical state by local
unitary operations, although $N_{G}^{p}$ is invariant under local
operations, $\sum\limits_{p=1}^{N}N_{K}^{p}$ varies under local unitary
operations.

Three qubit Greenberger-Horne-Zeilinger state 
\begin{equation}
\Psi _{GHZ}=\frac{1}{\sqrt{2}}\left( \left\vert 000\right\rangle +\left\vert
111\right\rangle \right) ,
\end{equation}%
is a maximally entangled state having pure tripartite entanglement. Consider
\ one parameter pure states

\begin{equation}
\Psi _{1}=\sqrt{q}\Psi _{GHZ}+\sqrt{\frac{\left( 1-q\right) }{2}}\left(
\left\vert 100\right\rangle +\left\vert 011\right\rangle \right) \text{, }%
\widehat{\rho }_{1}=\left\vert \Psi _{1}\right\rangle \left\langle \Psi
_{1}\right\vert ,  \label{26}
\end{equation}%
where $0\leq q\leq 1$. For these states, $N_{2}^{1}=0$ for all values of $q$
and $N_{G}^{1}=N_{3}^{1}$. Fig. (1) displays total negativities $%
N_{3}^{t}=\sum\limits_{p=1}^{3}N_{3}^{p}$ and $N_{2}^{t}=\sum%
\limits_{p=1}^{3}N_{2}^{p}$ \ versus $q$ for the states\ $\widehat{\rho }%
_{1} $. On applying a local rotation on qubit one, $N_{3}^{t}$ decreases,
reaches a minimum and then increases. The unitary rotation

\begin{equation*}
U_{1}\left( \theta _{1}\right) =\left[ 
\begin{array}{cc}
\cos \left( \theta _{1}\right) & \sin \left( \theta _{1}\right) \\ 
-\sin \left( \theta _{1}\right) & \cos \left( \theta _{1}\right)%
\end{array}%
\right]
\end{equation*}%
that minimizes $N_{3}^{t}$ for a given value of parameter $q$ is found,
numerically. The negativities $N_{2}^{t}(\theta _{1})$ and $N_{3}^{t}(\theta
_{1})$ of partial $2-$way and $3-$way transpose of state 
\begin{equation*}
\widehat{\rho }_{1}(\theta _{1})=\left\vert \Psi _{1}\left( \theta
_{1}\right) \right\rangle \left\langle \Psi _{1}\left( \theta _{1}\right)
\right\vert ,
\end{equation*}%
where%
\begin{equation*}
\Psi _{1}\left( \theta _{1}\right) =\left( U_{1}\left( \theta _{1}\right)
\otimes I_{2}\otimes I_{3}\right) \Psi _{1}
\end{equation*}%
are also displayed in Fig. (1).

The genuine tripartite entanglement, $E_{3}=\min ({E}_{3}^{1}(\theta
_{1}),{E}_{3}^{2}(\theta _{1}),{E}_{3}^{3}(\theta _{1}))$, is
displayed as a solid line plot in Fig. (2). Fig. (2) also shows $%
E_{2}^{2}(\theta _{1})$, which is a measure of bipartite entanglement of
qubits $2$ and $3$. The state $\widehat{\rho }_{1}(\theta _{1})$ has the
same degree of entanglement as the state $\widehat{\rho }_{1}$ but lower
values of $2-$way and $3-$way coherences. The canonical state corresponding
to state $\Psi _{1}$, found by the procedure given by Acin et al \cite%
{acin01} is a GHZ like state%
\begin{equation*}
\Psi _{1}^{\prime }=\sqrt{\frac{1-2q}{2}}\left\vert 000\right\rangle +\sqrt{%
2q(1-q)}\left\vert 100\right\rangle +\sqrt{\frac{1}{2}}\left\vert
111\right\rangle ,
\end{equation*}%
and is equal to state $\Psi _{1}\left( \theta _{1}\right) $, obtained
numerically.

In general, by successive unitary rotations of qubits one, two, and three,
one may obtain a state operator with minimum $2-$way and $3-$way coherences.
Minimization of $\sum\limits_{p=1}^{N}N_{K}^{p}$, by successive\ unitary
rotations of qubits, offers a method to reach the canonical state from a
given state of the $N$ qubit composite system. In most cases, a simple
computer program suffices to effect such a minimization, numerically.
Analytical proof and further discussion on using minimization of $K-$way
negativities to reach the canonical state, will be given elsewhere. 
\begin{figure}[t]
\centering \includegraphics[width=3.75in,height=5.0in,angle=-90]{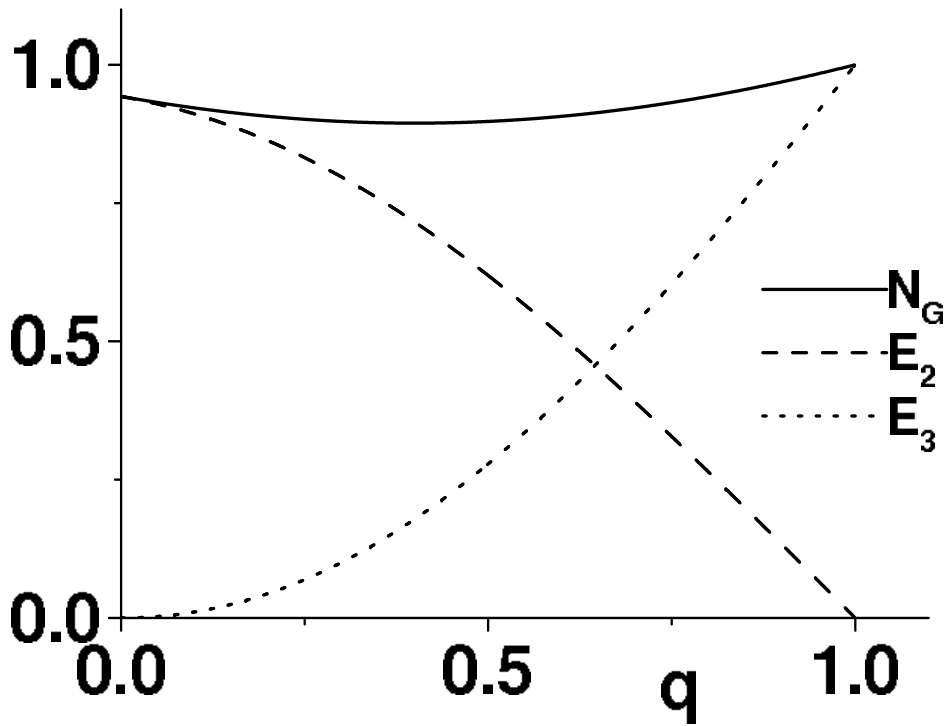}
\caption{ Global Negativity $N_{G}$, partial $2-$way negativity $E_{2}$, and
partial $3-$way negativity $E_{3}$ as function of parameter $q$ for the
state $\hat{\protect\rho}_{1}$.}
\label{fig3}
\end{figure}

\section{Three qubit GHZ and W-type states}

There exists a class of tripartite states akin to maximally entangled
W-state given by 
\begin{equation}
\Psi _{W}=\frac{1}{\sqrt{3}}\left( \left\vert 100\right\rangle +\left\vert
010\right\rangle +\left\vert 001\right\rangle \right) .  \label{24}
\end{equation}%
To illustrate the ideas presented in previous sections, we analyze the
entanglement of a set of single parameter pure states of the type

\begin{equation}
\Psi _{2}=\sqrt{q}\Psi _{GHZ}+\sqrt{1-q}\Psi _{W}\text{ , }\widehat{\rho }%
_{2}=\left\vert \Psi _{2}\right\rangle \left\langle \Psi _{2}\right\vert ,
\label{25}
\end{equation}%
for $0\leq q\leq 1.$

Fig. (3) displays $%
E_{2}^{1}=E_{2}^{2}=E_{2}^{3}=E_{2},E_{3}^{1}=E_{3}^{2}=E_{3}^{3}=E_{3}$,
and $N_{G}^{1}=N_{G}^{2}=N_{G}^{3}=N_{G}$ as a function of parameter $q$ for
a single qubit in state $\widehat{\rho }_{2}$. Global negativity indicates
that all three qubits are entangled as $q$ varies from zero to one. For $%
q=0, $ $E_{2}=0.94$, and $E_{3}=0,$ that is the system does not have genuine
tripartite entanglement. On the other hand for $q=1,$ the system has maximum
genuine tripartite entanglement as is evidenced by $E_{3}=1.0,$ and $E_{2}=0$%
. The measure of genuine tripartite entanglement is $E_{3}$ for $0\leq q\leq
1$. The W-state like tripartite entanglement is generated by $2-$way quantum
correlations.

\begin{figure}[t]
\centering \includegraphics[width=3.75in,height=5.0in,angle=-90]{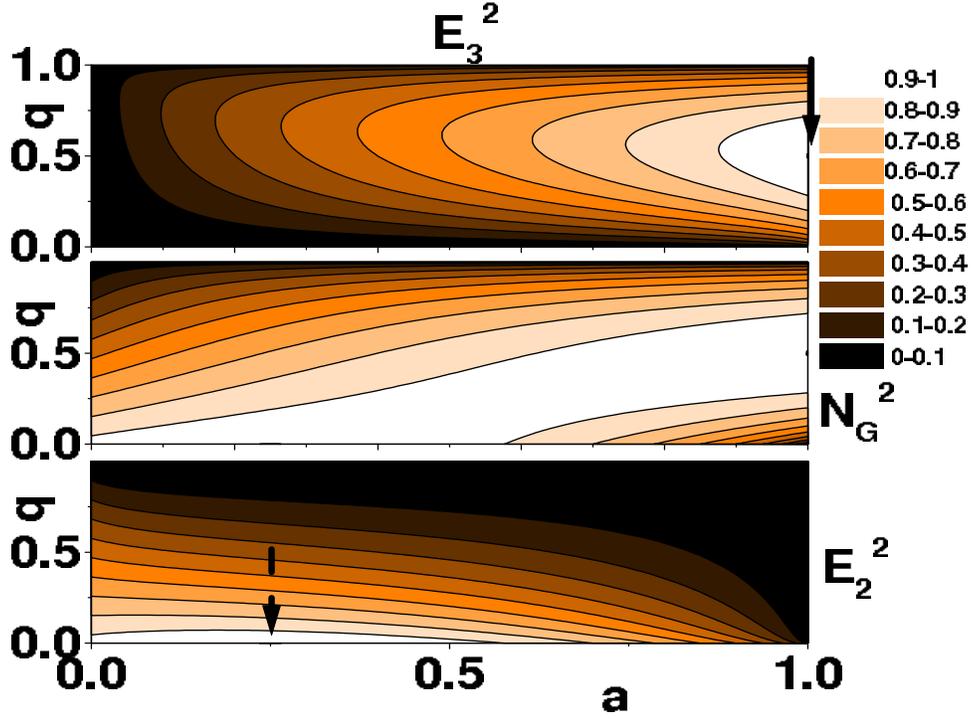}
\caption{Contour plots of genuine tripartite entanglement $E_{3}^{2}$, 
global Negativity $N_{G}^{2}$, and bipartite entanglement $%
E_{2}^{2}$, as a function of parameters $q$ and $a$ for two mode three boson 
state, $\hat{\protect\rho}_{3} $. Solid line arrow and dashed line arrow 
indicate the states with maximal
GHZ-type and maximal W-type entanglement, respectively. 
}
\label{fig4}
\end{figure}

\section{Three qubit two parameter canonical state}

To further illustrate the usage of $K-$way negativities, we consider a three
qubit two parameter canonical state%
\begin{equation}
\Psi _{q,a}=\sqrt{a}\left\vert 000\right\rangle +\sqrt{\left( 1-q\right)
\left( 1-a\right) }\Psi _{W}+\sqrt{q\left( 1-a\right) }\left\vert
111\right\rangle \text{, }\widehat{\rho }_{3}=\left\vert \Psi
_{q,a}\right\rangle \left\langle \Psi _{q,a}\right\vert \text{.}  \label{30}
\end{equation}%
Zheng et al. \cite{zhen05} have shown $\Psi _{q,a}$ to represent a two mode
three boson state and calculated the concurrence and tangle for the state.
Fig. (4) displays the contour plots of entanglement measures $E_{3}^{2}$, $N_{G}^{2}$, 
and $E_{2}^{2}$ for this state. For all values of parameters $a$ and $q$, genuine
tripartite entanglement is given by $E_{3}^{1}=E_{3}^{2}=E_{3}^{3}$
and partial $2-$way negativity resulting in bipartite and W-state like
entanglement is given by $E_{2}^{1}=E_{2}^{,2}=E_{2}^{3}$. The state with maximal
genuine tripartite entanglement and the state with maximal W-type
entanglement are indicated by solid line arrow and dashed line arrow, respectively.

\begin{figure}[t]
\centering
\par
\centering \includegraphics[width=3.75in,height=5.0in,angle=-90]{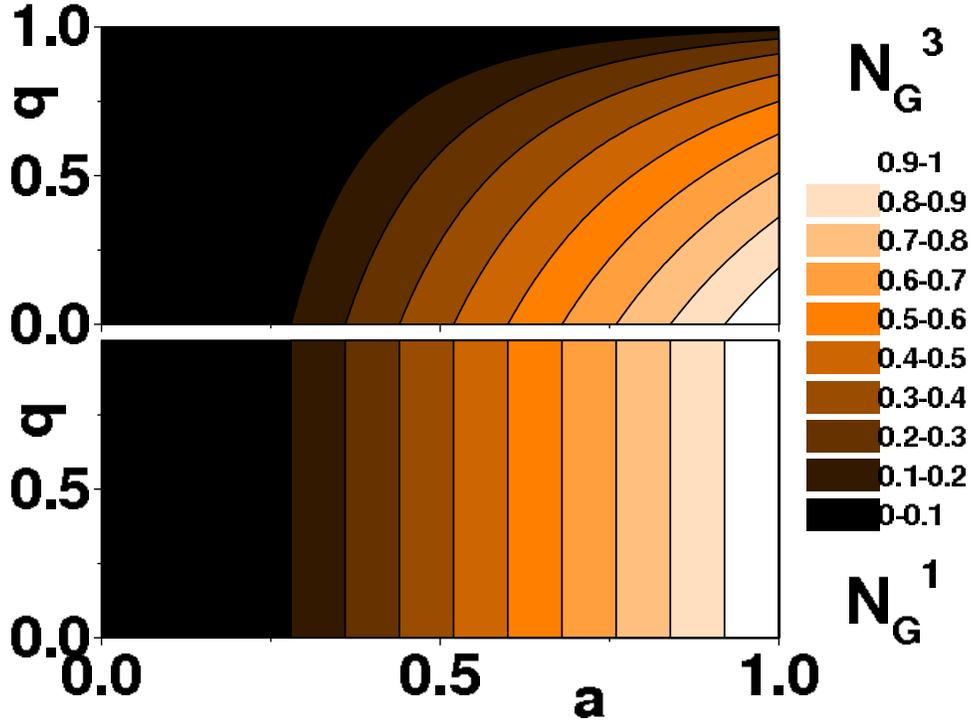}
\caption{Contour plots of Global Negativities $N_{G}^{3}$ and $%
N_{G}^{1}$ plotted as a function of parameters $q$ and $a$, for
the state $\hat{\protect\rho}_{4}$}
\label{fig5}
\end{figure}

\section{Three qubit noisy state}

Figs. (5) and (6) display entanglement measures for two parameter three
qubit noisy states of the form

\begin{equation}
\widehat{\rho }_{4}=a\left\vert \Psi _{q}\right\rangle \left\langle \Psi
_{q}\right\vert +(1-a)\frac{\widehat{I}_{8}}{8},
\end{equation}%
where

\begin{equation}
\Psi _{q}=\frac{1}{\sqrt{2}}\left[ \left\vert 000\right\rangle +\sqrt{1-q}%
\left\vert 111\right\rangle +\sqrt{q}\left\vert 110\right\rangle \right] ,
\label{32}
\end{equation}%
$0\leq q\leq 1$ and $0\leq a\leq 1.$ Global negativity $N_{G}^{1}=N_{G}^{2}$
for qubits one and two is independent of parameter $q$ and varies linearly
with $a$ as seen in contour plots of Fig. (5). For $a=0$ we have $\widehat{%
\rho }_{4}=\frac{\widehat{I}_{8}}{8}$ with $N_{G}=E_{3}={E}_{2}=0$%
 that is a separable mixed state. On the other hand for the
case $a=1,$ $q=0$ ( $\Psi _{0}=\Psi _{GHZ})$, we have 
$N_{G}=E_{3}=1.0$, and $E_{2}=0$ for all qubits. Global negativity values
indicate that all three sub-systems remain separable for $a\leq 0.2$. The
contour plots of entanglement measures $E_{2}^{1}=E_{2}^{2}$ and $E_{3}$ in 
Fig. (6) show the regions of maximum genuine tripartite
entanglement, bipartite entanglement and separability. As qubit three has
only tripartite entanglement $E_{2}^{3}$ being zero, the tripartite
entanglement measure $E_{3}$ is equal to $N_{G}^{3}$. For the
parameter values $a=1$, $q=1$, the system has maximum bipartite entanglement
but no tripartite entanglement as the state looks like 
\begin{equation*}
\left\vert \Psi _{q=1}\right\rangle =\left( \frac{\left\vert 00\right\rangle
+\left\vert 11\right\rangle }{\sqrt{2}}\right) \left\vert 0\right\rangle .
\end{equation*}

\begin{figure}[t]
\centering \includegraphics[width=3.75in,height=5.0in,angle=-90]{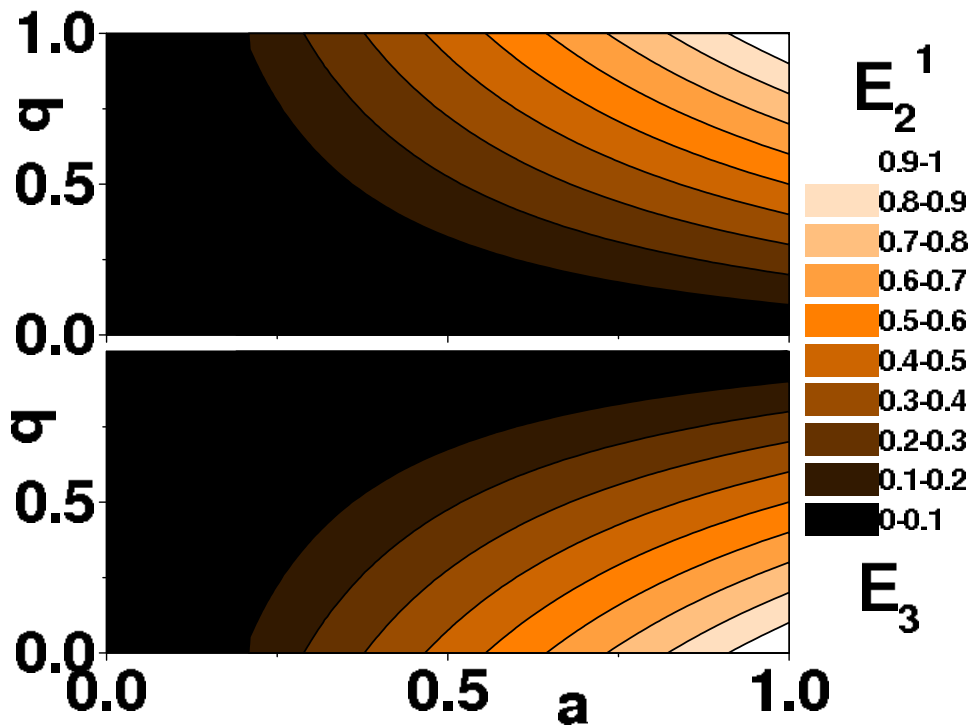}
\caption{Contour plots of bipartite entanglement measure $E_{2}^{1}$ %
and genuine tripartite entanglement $E_{3}$ plotted as a function of
parameters $q$ and $a$, for the state $\hat{\protect\rho}_{4}$.}
\label{fig6}
\end{figure}

\section{Conclusions}

We have defined global \cite{vida02} and $K-$way negativities calculated
from global and $K-$way partial transposes, respectively, of an $N-$partite
state operator. For a given partition of an N-partite system, global
negativity measures overall entanglement of parties. The $K-$way
negativities for $2\leq K\leq $ N, on the other hand, provide\ a measure of $%
K-$way coherences of the system. Global negativity with respect to a
subsystem can be written as a sum of partial $K-$way negativities. We
conjecture that the partial $K-$way negativities provide an entanglement
measure for $N-$partite canonical states. For canonical states, the
coherences have their minimum value as such definite relations exist between
the negativities and $K-$partite entanglement of these states. We have
applied these ideas to one and two parameter three qubit states. For a three
qubit system a combined analysis of $\ 2-$way, $3-$way and global
negativities with respect to all subsystems is shown to provide distinct
measures of genuine tripartite, W-type, and bipartite entanglement of the
composite system. Entanglement is invariant with respect to local unitary
operations, whereas, coherences are not so. Extension to qutrits and
application to systems with more than three parties, should be possible.

\section{Acknowledgements}

Financial support from National Council for Scientific and Technological
Development (CNPq), Brazil and State University of Londrina, (Faep-UEL),
Brazil is acknowledged.

\end{document}